\documentclass[%
 aapm,
 mph,%
 amsmath,amssymb,
 reprint,%
 author-numerical,%
]{revtex4-2}

\usepackage{comment}
\usepackage{graphicx}
\usepackage{dcolumn}
\usepackage{bm}
\usepackage{xcolor}

\begin{document}

\preprint{AAPM/123-QED}

\title[Trajectory-Dependent Electronic Energy Losses in Ion Range Simulations]{Trajectory-Dependent Electronic Energy Losses in Ion Range Simulations} 
\author{Glen P. Kiely}
\email{glen.kiely@aalto.fi.}
\author{Bruno Semião}%
\author{Evgeniia Ponomareva}%
\author{Rafael Nuñez-Palacio}%
\author{Unna Arpiainen}%
\author{Andrea E. Sand}%
\affiliation{ 
Aalto University, Finland 
}

\date{\today}

\begin{abstract}

The energy losses of energetic ions in materials depend on both nuclear and electronic interactions. In channeling geometries, the stopping effect of these interactions can be highly reduced, resulting in deeper ion penetration. Comprehensive, trajectory-dependent models for ion-material interactions are therefore crucial for the accurate prediction of ion range profiles. We present the implementation of a recent electron density-dependent energy-loss model in the efficient molecular dynamics-based MDRANGE code. The model captures \textit{ab initio} electron dynamics using a parametrized ion energy loss function, based on calculations for explicit trajectories using real-time time-dependent density functional theory. We demonstrate the efficient simulation of trajectory-dependent ion range profiles with this comprehensive model for electronic energy losses. Our results indicate that accurate trajectory-dependent ion range profiles can be simulated using well-fitted parametrizations of this model. This method offers a unique tool for validation of the fitted energy-loss functions using energetic ion ranges, which can be measured experimentally but are beyond the capability of full MD simulations due to the computational expense. 
\end{abstract}

\maketitle

\section{\label{sec:intro} Introduction}

Ion irradiation of materials is encountered in a wide range of research and industrial applications, including semiconductor doping \cite{Williams1998}, ion-beam lithography \cite{lithography_review}, and in the study of radiation-induced damage in nuclear materials \cite{ranges_damage}. Understanding and predicting the effects of ion irradiation is crucial for optimizing material modification processes and for assessing irradiation-induced material damage \cite{was2017}. Computational modeling provides a powerful and cost-effective approach for simulating ion-matter interactions and insight into these radiation effects. 

The predictive capability of computational models depends on the accurate description of interactions between energetic ions and target materials. These interactions lead to a reduction in the kinetic energy of the ion, quantified by the stopping power: \(-\frac{dE}{dx} = S_e + S_n\), where \(S_e\) is the electronic stopping power (ESP) representing the energy transfer from the energetic ion to target electrons; and \(S_n\) is the nuclear stopping power (NSP), which arises from interactions between energetic ions and target atomic nuclei \cite{schmidt2013}.

Ion irradiation processes can be modeled using molecular dynamics (MD) simulations or Monte Carlo (MC) methods based on the binary collision approximation (BCA). MD simulations dynamically resolve atomic-scale interactions through Newtonian mechanics, enabling high-accuracy modeling of NSP using interatomic potentials \cite{teunissen2023, sillanpaa2001}. In contrast, MC-based BCA methods model ion interactions and energy losses statistically, sacrificing atomic-scale accuracy in favor of computational efficiency, allowing for the simulation of many irradiation events. MDRANGE is an MD-based simulation code that was specifically developed to overcome the computational challenge of using MD to predict ion range profiles \cite{nordlund_1994_mdrange}, and offers an efficiency comparable to that of BCA methods with the atomic-scale accuracy of MD. 

The ESP is widely modeled as a scalar velocity-dependent retarding force according to Lindhard theory \cite{lindhard1965}, which has been expanded to improve accuracy across different energy regimes. 
Values for the ESP may be obtained numerically or from computational programs such as PSTAR for protons \cite{PSTAR}, MSTAR for heavier ions (3Li to 18Ar) \cite{MSTAR}, or SRIM (The Stopping and Range of Ions in Matter) for all ions. \cite{SRIM} These programs combine theoretical frameworks, scaling laws, and empirical data to provide predictions for the average ESP of ions in materials. 

However, projectile ions can travel along trajectories in which they experience lower than average atomic and electronic densities. This effect is known as channeling and occurs when the ions are directed by interatomic potentials along the principal crystal axes (axial) or between planes or rows of atoms (planar) \cite{lindhard1965, nordlund_2016_channel}. In these channeling geometries, projectile dynamics are affected less by lattice nuclei, and they lose their kinetic energy mainly through electronic interactions. Ions can also experience varying electronic interactions in these geometries because of the lower electron densities they encounter. These effects lead to a reduction in the total stopping power on average \cite{sigmund1998} and result in higher ion ranges (penetration depths) in channeling, compared to non-channeling, geometries \cite{nordlund_2016_channel}. 

Real-time time-dependent density functional theory (rt-TDDFT) has been utilized to simulate the temporal evolution of quantum electronic interactions of energetic ions in materials \cite{Nunez2025, lee_2020_tddft, lim_2016, tamm_2019_eph1, teunissen2023, jarrin_2021, ullah2018}.  While rt-TDDFT can provide an accurate, first-principles description of electronic energy losses, its computational cost is prohibitive for large-scale ion irradiation simulations. In the approach outlined by Correa et al. \cite{correa_2018_esfp}, a trajectory- and energy-dependent scalar ESP can be obtained by analyzing the ion's rate of energy loss as a function of its displacement in short trajectories and thus utilized in larger-scale simulations. 

Recently, Tamm et al. \cite{tamm_2018_eph2} developed a model that uses rt-TDDFT energy loss data to parametrize a coupling function capturing the effects of electronic interactions on a projectile ion, dependent on its local environment, in a two-temperature molecular dynamics (TTMD) framework. This approach enables modeling of a trajectory-dependent ESP with \textit{ab initio} accuracy in large-scale ion irradiation simulations. The model is implemented for the MD code LAMMPS \cite{LAMMPS} in the USER-EPH plugin \cite{tamm_2019_eph3}. 

In this work, we present the implementation of the model developed by Tamm et al. \cite{tamm_2018_eph2}, which we hereby refer to as the unified two-temperature model (UTTM), in the MD-based code MDRANGE. This new implementation allows modeling ion ranges over micrometers in depth with \textit{ab initio} based trajectory-dependent electronic energy losses, with statistics from tens of thousands of ions, which is inaccessible by full MD. Through comparison to other models in suitably chosen geometries, and to experimental measurements where such are available, this method also provides a tool for validating the parametrized coupling functions of the UTTM model. We investigate the performance of the UTTM model implementation in MDRANGE through simulations of ion range profiles in a variety of materials along both channeling and non-channeling trajectories. 

We demonstrate the trajectory dependence of the predicted ion ranges using the UTTM model in MDRANGE, and compare the results with simulations carried out using other ESP models. Overall, we show that the UTTM model captures trajectory-dependent electronic energy losses and provides efficient predictions of ion range profiles in arbitrary materials along different crystal orientations. 

\section{\label{sec:elec_stop} Electronic Stopping Power Models}

For any ESP model, the simulated ion range profiles vary along different crystal orientations, due to the different atomic environments encountered \cite{nordlund_2016_channel}. Adjusting the ESP model further modifies these predictions. In this work, we compare trajectory-dependent ion range profile predictions in a variety of materials using different models for the ESP. We categorize two distinct ESP model types: {(1) scalar models, whereby a velocity-dependent retarding force is applied to the projectile ion without accounting for the ion's local environment, and (2) trajectory-dependent models, such as the UTTM model, which explicitly consider the ion's local environment in the calculation of the retarding force on the projectile.}

\subsection{\label{subsubsec:scalar_models} Scalar Electronic Stopping Power}

{The scalar model approach treats the ESP as a velocity-dependent friction force that decelerates the ion. In an MD framework, the force on an atom is defined by:}

\begin{equation} \label{eq:full_EPH}
\mathbf{f}_I = -\nabla_I U - \beta_I\left(v_I\right)\mathbf{v}_I
\end{equation}

{where $U$ is the interatomic potential, and $\beta(v)$ is the scalar velocity-dependent ESP friction force.} {In this work, we employ values for the scalar ESP} calculated by SRIM \cite{SRIM}, which is widely utilized in radiation damage and ion range studies. The ESP is calculated by SRIM semi-empirically, and is based on theoretical models fit to experimental data where it is available. This experimental data is scarce, particularly for heavier ions at low velocities. In these cases, the calculated ESP is extrapolated to the low-velocity regime, introducing uncertainties in modeling the electronic energy losses of slow ions \cite{PREPRINT_Nickel}. 

Moreover, a significant limitation of SRIM is that it does not account for the ordered lattice or electronic structure of materials. Instead, the ESP calculated by SRIM is generally valid for random ion trajectories in target materials. For trajectories where the local electronic and nuclear environments differ significantly from the average random trajectory, such as along channeling directions, electronic energy losses are not accurately modeled \cite{sand2019}. 

Unlike the semi-empirical approach, rt-TDDFT directly resolves the energy exchanged between projectile ions and the target electrons, which allows for a velocity- and trajectory-dependent description of electronic energy losses. A scalar ESP force on the ion can be calculated for chosen trajectories as outlined by Correa et al. \cite{correa_2018_esfp}. While this provides higher fidelity in modeling the ESP of ions which travel along channels, it does not account for the local ion environment. Consequently, it does not accurately model electronic energy losses for projectiles which are either not significantly channeled, or have been scattered out of the channel. Nevertheless, the electronic energy losses for strongly channeled ions are captured correctly in range simulations employing a constant rt-TDDFT value for the ESP. Hence, this approach yields accurate predictions of the maximum ion range $R_{max}$  \cite{sand2019}, which is typically determined in ion range experiments \cite{eriksson1967}. 

\subsection{\label{subsec:local_models} Trajectory-Dependent Electronic Stopping Power}

{Accurately capturing the ESP requires a model which accounts for variations in the local electronic environment encountered along ion trajectories.}
The {trajectory-dependent} UTTM model is based on Langevin dynamics, in which particles in a system experience frictional forces and corresponding random forces via the fluctuation-dissipation theorem. The UTTM model modifies the traditional Langevin dynamics by replacing forces on individual atoms with correlated many-body forces, and a tensorial frictional force describes the ESP acting on the atoms \cite{tamm_2018_eph2}. The force on an atom is defined by:

\begin{equation} \label{eq:full_EPH}
\mathbf{f}_I = -\nabla_I U - \underbrace{\sum_J \mathbf{B}_{IJ} \mathbf{v}_J}_{\boldsymbol{\sigma}_I} + \underbrace{\sum_J \mathbf{W}_{IJ} \boldsymbol{\xi}_J}_{\boldsymbol{\eta}_I}
\end{equation}


where $\bm{\sigma}_I$ is the tensorial friction applied to each atom, which describes the ESP of the projectile ion, and is proportional to the relative velocities of the neighborhood atoms. The correlated random fluctuations are described by $\boldsymbol{\eta}_I$. Each $3 \times 3$ tensor $\bm{B}_{IJ} = \sum_K \bm{W}_{IK} \bm{W}^T_{JK}$ is positive-definite, and $\bm{W}_{IJ}$ is given by:

\begin{equation} \label{frictioncalceq}
    \bm{W}_{IJ} =
    \begin{cases}
        -\alpha_J \frac{\rho_I \left( r_{IJ} \right)}{\bar{\rho}_J} \bm{e}_{IJ} \bm{e}_{IJ}, & (I \neq J) \\
        \alpha_I \sum_{K \neq I} \frac{\rho_K \left( r_{IK} \right)}{\bar{\rho}_I} \bm{e}_{IK} \bm{e}_{IK}, & (I = J)
    \end{cases}
\end{equation}

{where $\rho_I(r_{IJ})$ is the contribution to the electron density of atom $I$ at a distance $r_{IJ}$, and $\bar{\rho}_J = \sum_{I \ne J} \rho_I(r_{IJ})$ is the total local electron density experienced by atom $J$. The vector $\bm{e}_{IJ}$ is the unit vector joining atoms $J$ and $I$.
The coupling function $\alpha_J = \alpha(\bar{\rho}_J)$ is dependent on the total electron density $\bar{\rho}_J$ experienced by atom $J$.} The correlated random forces are determined by a set of independent white noise Gaussian variables $\langle \xi_I(t)\rangle_I$ given by:

\begin{equation} \label{eq:rand_forces}
\langle \xi_I(t) \xi_J(t') \rangle = 2 k_B T_e \delta(t - t') \delta_{IJ}
\end{equation}

where $T_e$ is the temperature of the electronic system \cite{tamm_2018_eph2}. The $\alpha (\bar{\rho})$ coupling component is determined from \textit{ab initio} rt-TDDFT data. However, the accuracy of this model depends on the ability of the parameterized  single-valued coupling function to capture the complex rt-TDDFT data, and determining a suitable parameterization can be challenging.  

\section{\label{sec:overall_methods} Simulation Methodology}

\subsection{\label{sec:approximations} MDRANGE Approximations}

MDRANGE enables the efficient simulation of ion range profiles through two key approximations. These approximations reduce computational complexity while maintaining accurate modeling of projectile ion dynamics\cite{nordlund_1994_mdrange}:

1) The recoil interaction approximation (RIA): interatomic interactions between the material's lattice nuclei are neglected. Instead, only the interactions between the projectile ion and the lattice nuclei are explicitly modeled, significantly simplifying the simulation complexity while capturing the essential projectile ion dynamics to high accuracy \cite{nordlund_1994_mdrange}. 

2) Localized simulation domain: Unlike traditional MD, which models the entire system relevant to the process being studied, MDRANGE simulates a small, localized region of the target material around the ion trajectory. Lattice nuclei are generated on-the-fly as the ion enters their vicinity, allowing for the computationally efficient simulation of long ion trajectories. 

\subsection{\label{sec:methods} Ion Range Profile Calculation}

Projectile ions are initialized at randomly distributed positions near the surface of the target. 
The target materials are thermalized with displacements determined by the Debye model, as detailed in Ref. \cite{debye_disp}. Nuclear interactions between projectile ions and lattice atoms are modeled using the Ziegler–Biersack–Littmark (ZBL) repulsive potential \cite{ZBL_book}. Each ion trajectory is simulated until the kinetic energy of the projectile reaches 1 eV. {In accordance with ion range experiments \cite{eriksson1967}, we present the ion range profiles as integrated distributions which show the fraction of implanted ions that reached a given range.}

\subsection{\label{sec:EPH_in_MDRANGE} Implementation of Trajectory-Dependent Electronic Energy Losses in MDRANGE}

{We have implemented the UTTM model in MDRANGE, to incorporate trajectory-dependent electronic energy losses in ion range simulations. The UTTM model was originally developed for use in full MD and accounts for correlated motion of lattice atoms, hence the calculations of the friction components include the positions and velocities of neighboring atoms in a large region around the projectile and their effects on each other.}

\begin{figure} 
\centering 
\includegraphics[width=0.49\textwidth]{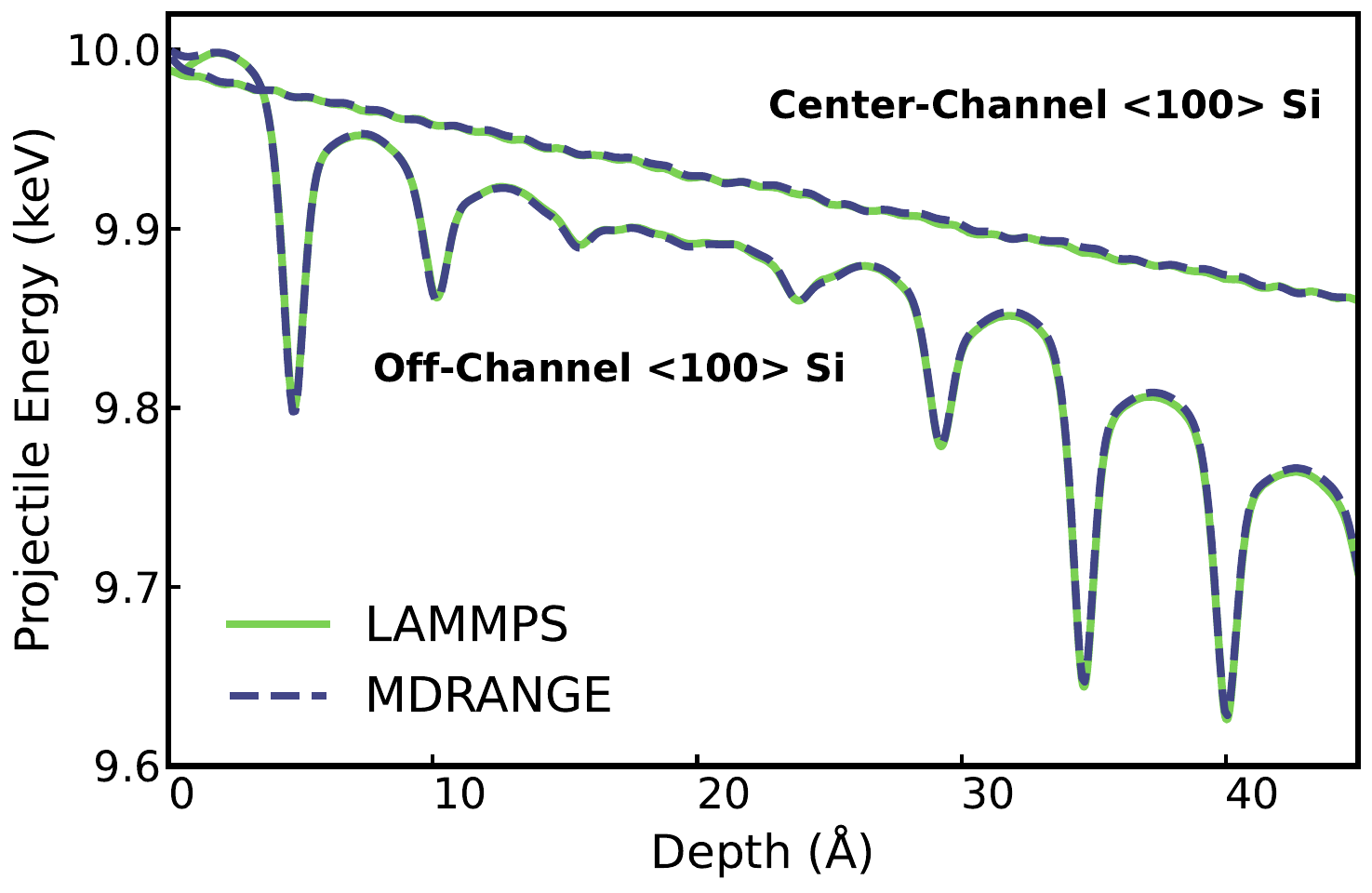} 
\caption{\label{fig:lammpsvsmdrange} 
Comparison of the rate of energy loss of 10 keV Si ions traveling along center-channel and off-channel \(\langle 100 \rangle\) trajectories in an Si lattice, {simulated using LAMMPS and MDRANGE. The ESP was modeled using the UTTM model with a quadratic $\alpha (\bar{\rho})$ function developed by Jarrin et al. \cite{jarrin_2021}}. The target was initialized at 0 K without thermal displacements. In the off-channel trajectory, the projectile ion was shifted toward a neighboring lattice atom by $50\%$ of the distance between them in the center-channel trajectory. 
} 
\end{figure}

{Analogous to the RIA, in our implementation of the UTTM model in MDRANGE we apply the UTTM tensorial forces only to the projectile ion's dynamics, neglecting electronic effects on the lattice nuclei. This simplifies the treatment of the arising forces while preserving the accuracy in modeling ion trajectories. In Fig. \ref{fig:lammpsvsmdrange}, we demonstrate that this approximation does not affect the calculated energy losses through comparison with the USER-EPH package in LAMMPS \cite{tamm_2018_eph2,tamm_2019_eph3}, which offers full MD treatment of the projectile ion and lattice atoms. We compare portions of single 10 keV Si ion trajectories directed along the \(\langle 100 \rangle\) channel in Si, both in a center channel trajectory and for a trajectory shifted off-center along the direction of the channel. For these short trajectories, the maximum pointwise difference of MDRANGE relative to LAMMPS is less than 0.03\%.}

\begin{figure}
\includegraphics[width=0.49\textwidth]{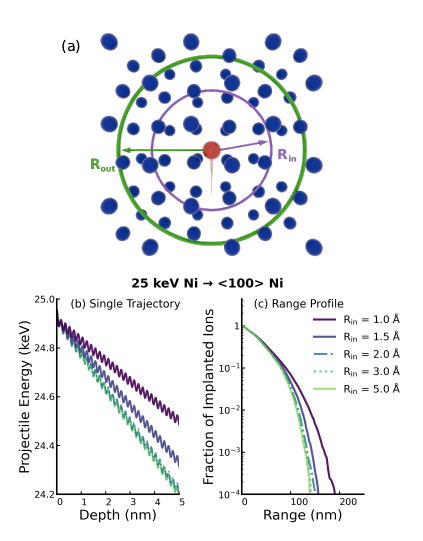}
\caption{\label{fig:fig_innerrad} (a) Illustration of a kinetic ion in a lattice, annotated with the radii used for the calculation of the ion's neighbors. The outer radius $R_{out}$ portrays the cutoff of the UTTM parametrization $\alpha(\bar{\rho})$. The inner radius $R_{in}$ portrays the user-defined cutoff radius implemented in our approximation for the calculation of full friction components of the UTTM model in MDRANGE. (b) The energy loss, as a function of distance traveled, of a single 25 keV Ni ion in the \(\langle 100 \rangle\) center channel in Ni {without thermal displacements.} (c) {Calculated integral} range profile of 25 keV Ni ions along the \(\langle 100 \rangle\) channeling direction in Ni, {thermalized at 300 K,} for varying $R_{in}$. {Ranges were calculated using the method described in Sec. \ref{sec:methods} for 50,000 ion implants.} We employed the Ni $\alpha(\bar{\rho})$ parameterization {developed by Caro et al. \cite{tamm_2019_eph1}, which} has a cutoff of $5.0$~Å. Curves for $R_{in} = 3.0$ and $5.0$~Å coincide for the energy loss and range profile.} 
\end{figure}

Because target lattice atoms do not interact with each other due to the RIA, they move insignificantly during the time that the projectile passes the region, and they can be treated in a mean-field approximation. To further optimize the UTTM model calculations in MDRANGE, and retain the small size of the lattice generated on-the-fly, we therefore defined an \textquotedblleft inner radius\textquotedblright{} $R_{in}$ that includes a subset of the projectile ion's closest neighbors. We illustrate this approximation in Fig. \ref{fig:fig_innerrad}a. For neighbors within $R_{in}$, we account for their interactions with all their corresponding neighbors in the friction terms explicitly, including the projectile ion. For lattice nuclei outside $R_{in}$, but within the electron density cutoff $R_{out}$ for the UTTM model parametrization, we consider only the direct effect of the projectile ion, combined with a precalculated average contribution from other lattice atoms. 

The relevant averages for the calculation of the friction force, described in Eq. \ref{frictioncalceq}, are evaluated in a thermalized lattice undisturbed by the projectile. Atoms in the center of the simulation cell are chosen in order to include all necessary neighbor contributions. The average influence of $\mathbf{W}_{IJ}$ on the local velocity field and the average electron density are calculated: 

\begin{equation}
\begin{aligned}
\langle \mathbf{W}\mathbf{v} \rangle &= \frac{1}{N_\text{center}} \sum_{I \in \text{center}} \sum_{J} \mathbf{W}_{IJ}^{T} \mathbf{v}_J, \\
\langle \bar{\rho} \rangle &= \frac{1}{N_\text{center}} \sum_{I \in \text{center}} \bar{\rho}_I
\end{aligned}
\end{equation}

where $N_\text{center}$ is the number of lattice atoms included in the averaging. This approach increases the efficiency of the computation of the UTTM model and also eliminates the need to simulate a large lattice to resolve the interactions of distant neighbors. This maintains the necessary efficiency for the simulation of thousands of ion trajectories while preserving the precision of the modeled electronic energy losses of the projectile ions. 

Figure \ref{fig:fig_innerrad}b shows that the energy losses calculated by the UTTM model for a single Ni projectile ion trajectory along the center \(\langle 100 \rangle\) channel are in complete agreement with the full UTTM model calculation ($R_{in} = R_{out} = 5.0$ Å for the Ni parametrization) when $R_{in}$ is sufficiently large ($R_{in} \geq 3.0$ Å in this case). As $R_{in}$ decreases, the energy losses of individual ions are progressively underestimated, due to the incomplete account of neighbor contributions. Consequently, the reduced ESP impacts the predicted integrated range profile, as shown in Fig. \ref{fig:fig_innerrad}c. 

\begin{figure}
\includegraphics[width=0.49\textwidth]{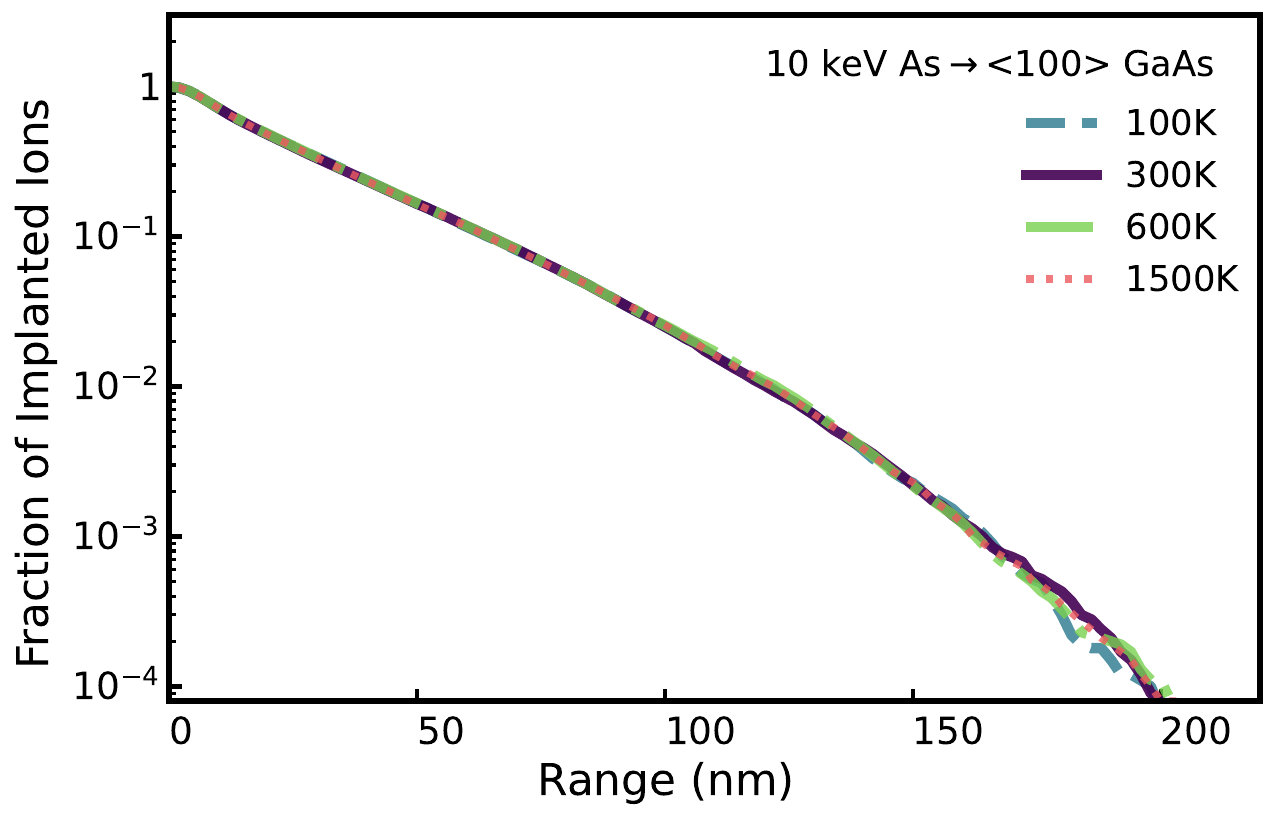}
\caption{\label{fig:electemp} Integral range profiles of 10 keV As ions in GaAs along the \(\langle 100 \rangle\) principal channeling direction using different constant values for the temperature of the electronic system. {The atomic system was thermalized to 300 K in all cases.} The range profiles were calculated {using the method described in Sec. \ref{sec:methods} for 100,000 ion implants}, with the UTTM model for the ESP. The GaAs coupling function used here for the UTTM model was developed by Teunissen et al. \cite{teunissen2023}. } 
\end{figure}

{The electronic temperature $T_e$ which appears in Equation \ref{eq:rand_forces} is not solved locally in MDRANGE, and instead we set a constant temperature equal to the initial material temperature.} We demonstrate in Fig. \ref{fig:electemp} that this does not significantly affect the range profile predictions in an exemplary case of 10 keV As ions implanted along the \(\langle 100 \rangle\) principal channeling direction in crystalline GaAs. The predicted range profiles do not show large discrepancies with different constant electronic system temperature values.

\begin{figure}
\includegraphics[width=0.49\textwidth]{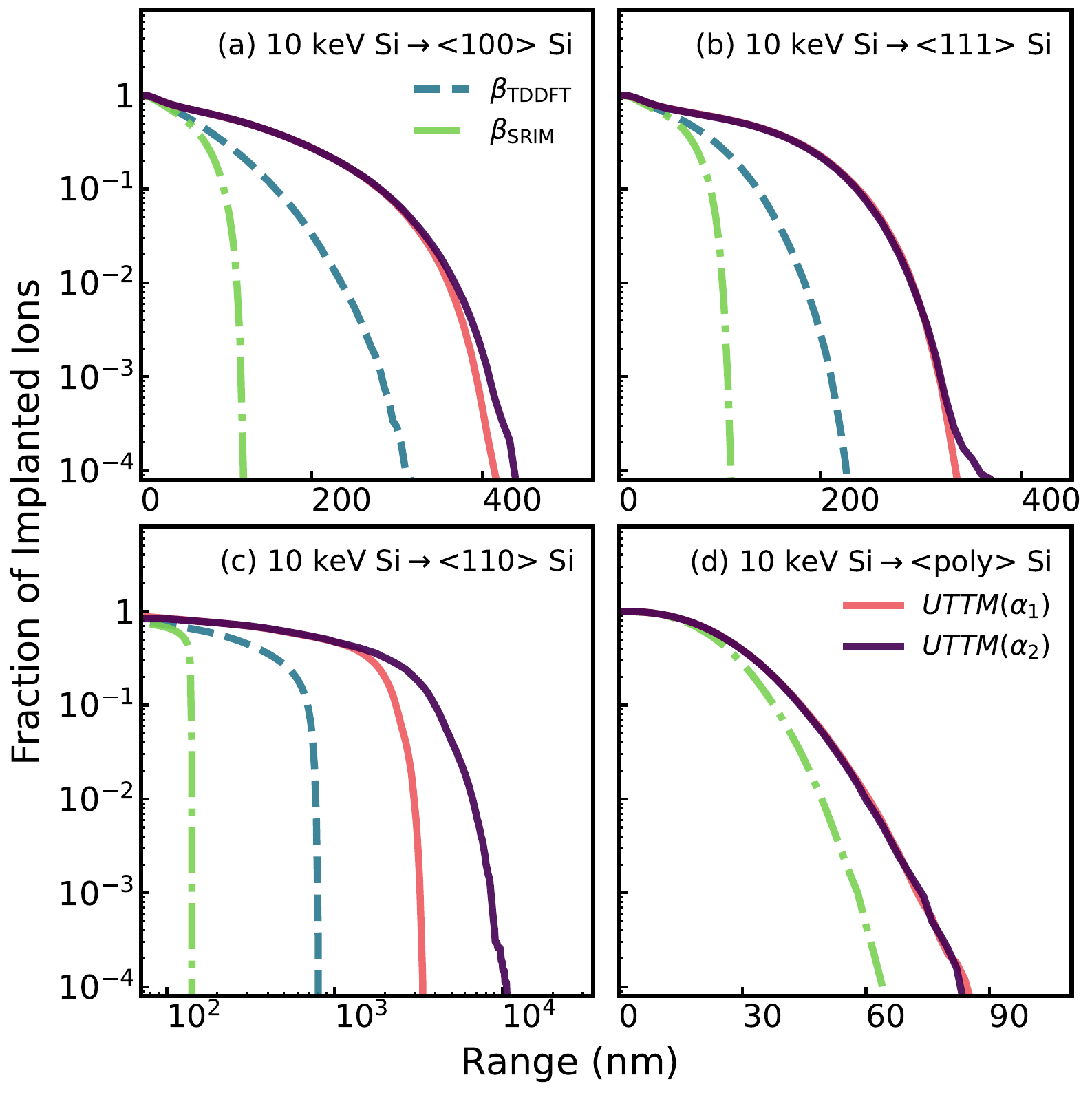}

\caption{\label{fig:Si_ranges} Calculated integral ion range profiles of 10 keV Si ions in diamond Si along the (a) \(\langle 100 \rangle\), (b) \(\langle 111 \rangle\), and (c) \(\langle 110 \rangle\) channeling directions, and (d) along a random direction, using different models for the ESP. Each channeling plot contains the range profiles predicted using the UTTM model with two parametrizations, both produced by Jarrin et al. \cite{jarrin_2021}, {which differ in the functional form of the $\alpha (\bar{\rho})$ coupling function.} Range profile predictions using the {scalar} ESP calculated using SRIM are plotted for all directions, and using scalar TDDFT calculated by Nuñez-Palacio et al. \cite{Nunez2025} for the channeling directions. In the \(\langle 100 \rangle\) and \(\langle 110 \rangle\) directions, the stopping corresponds to the center-channel, and in the \(\langle 111 \rangle\) direction it corresponds to the \textquotedblleft half-center\textquotedblright{} channel. Note the different axes scales, particularly the logarithmic x-axis scale in (c). 
}
\end{figure}

\section{\label{sec:results} Ion Range Profiles}

The UTTM model has been parameterized for several materials, including silicon (Si) by Jarrin et al. \cite{jarrin_2021}, nickel (Ni) by Caro et al.\cite{tamm_2019_eph1}, and gallium arsenide (GaAs) by Teunnisen et al. \cite{teunissen2023}. Here, {we select these materials based on the availability of these parametrizations and employ them} to calculate ion range profiles for self-ion irradiation in these materials, covering different projectile energies and crystallographic trajectories. 
{Using the method described in Sec. \ref{sec:methods}, we implant 100,000 projectile ions in targets thermalized at 300 K.} In the exceptional cases of implantation along the \(\langle 110 \rangle\) channeling direction in both Si and GaAs using the UTTM model, 25,000 projectiles were implanted due to large implantation depths. 

{We compare ion range profiles obtained using the UTTM model with those obtained using a scalar friction ESP model, employing stopping values from SRIM and from rt-TDDFT calculations. Simulations were performed along a variety of channeling directions, where ions experience larger impact parameters on average. Along channeling trajectories, ion range simulations using a rt-TDDFT-calculated ESP yield accurate predictions for the maximum range $R_{max}$, whereas a SRIM-calculated ESP fails to reliably describe electronic energy losses \cite{sand2019}.} 

{We also simulate ranges along a random direction, where close collisions play a significant role in the ions energy dissipation. Because the SRIM-calculated ESP captures the average electronic energy loss along random trajectories \cite{sand2019}, we expect the ion-range profiles simulated using SRIM to agree with those obtained from the UTTM model in this case.} Although in principle a random trajectory can be modeled by choosing a nonchanneling direction for the ion, this choice is dependent on the lattice structure of the material. Furthermore, many directions have been found to lead to some degree of channeling \cite{nordlund_2016_channel}. For consistency across different lattice structures, we model random trajectories by generating a polycrystalline target with an extremely small grain size of $50 \pm 5$ Å in random relative orientations, using the functionality available in the standard MDRANGE code. This approach ensures that the projectile ions cannot be significantly channeled. 

\subsection{\label{subsec:Si_ranges}  Silicon}

The Si diamond lattice was generated with the experimentally determined lattice constant of $a_0 = 5.43$ Å \cite{exp_lat_const}.
Figure \ref{fig:Si_ranges} illustrates the predicted range profiles of 10 keV Si projectiles in Si along low-index channeling directions and in random geometry calculated with the UTTM model using two different parameterizations developed by Jarrin et al. \cite{jarrin_2021}, which are characterized by quadratic and constant functional forms. For channeling trajectories along the \(\langle 100 \rangle\), \(\langle 111 \rangle\), and \(\langle 110 \rangle\) crystal axes, we additionally simulated range profiles using constant ESP values calculated with rt-TDDFT by Nuñez-Palacio et al. \cite{Nunez2025}. 

In each channeling direction, Si ions penetrate deeper when using the rt-TDDFT calculated ESP rather than that from SRIM. The predicted range profiles reveal a strong dependence on crystallographic orientation; the ion penetration depths are an order of magnitude larger along the \(\langle 110 \rangle\) channel than along other directions. This can be partly attributed to different atomic environments \cite{nordlund_2016_channel}. However, differences in predictions along the same directions using different models for the ESP further highlight the importance of electronic energy losses on ion ranges. Predictions employing the SRIM-calculated ESP resulted in the shortest ranges in every case, which is in agreement with its larger stopping values relative to rt-TDDFT calculations \cite{Nunez2025}. Along random directions (Fig. \ref{fig:Si_ranges}d), the range profile calculated with the trajectory-dependent UTTM model is in close agreement to the range obtained using SRIM with both parametrizations. 

\subsection{\label{subsec:Ni_ranges} Nickel}

\begin{figure}
\includegraphics[width=0.49\textwidth]{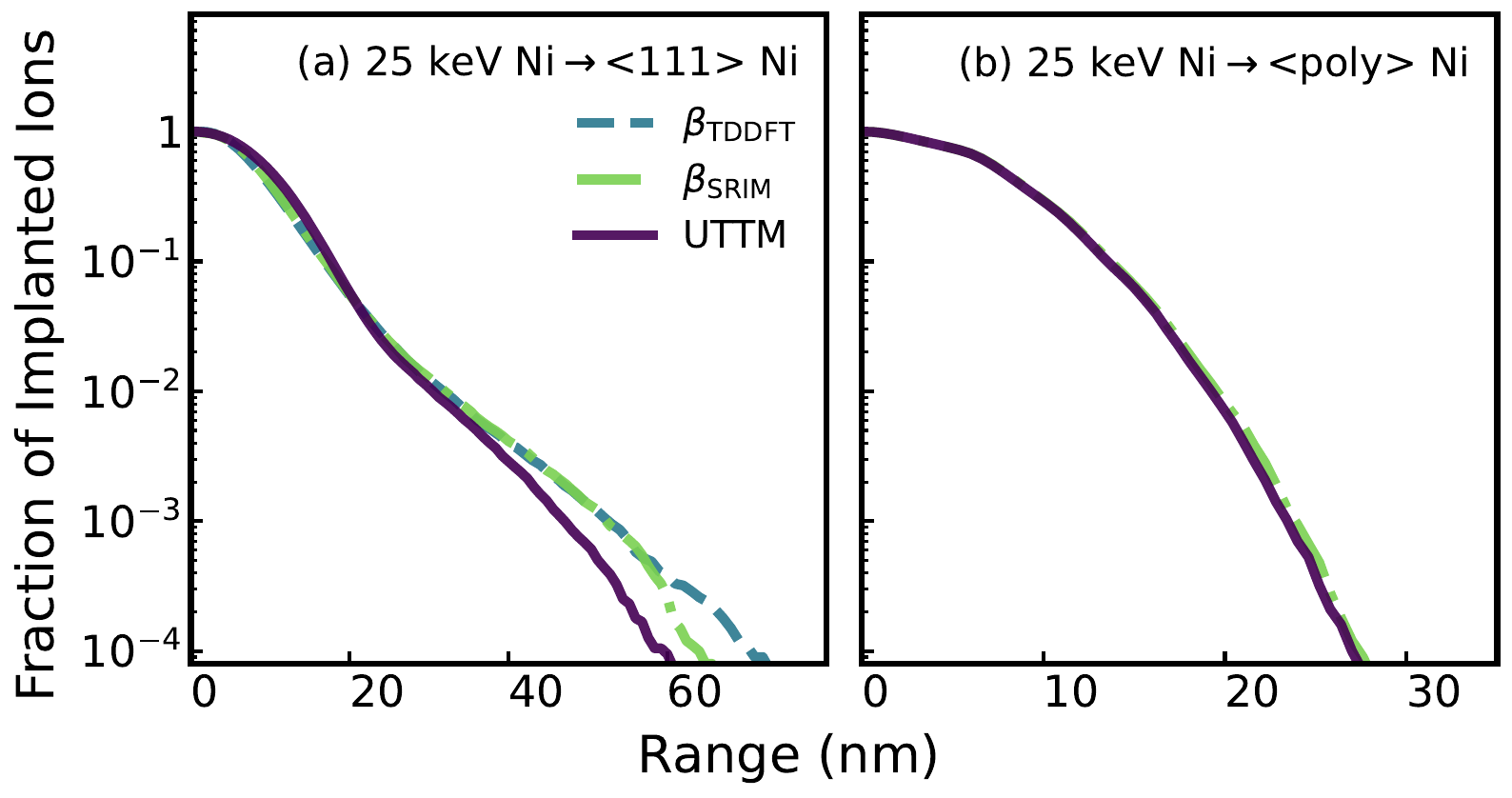}
\caption{\label{fig:Ni_prof} Calculated integral ion range profiles of 25 keV Ni ions in FCC Ni along the \(\langle 111 \rangle\) channel and a random direction, calculated using different models for the ESP. The \(\langle 111 \rangle\) channeling plot contains the predicted ion range profile using the UTTM model with the parametrization of \textit{ab initio} TDDFT data produced by Caro et al. \cite{tamm_2019_eph1}, and the scalar TDDFT data calculated by Ullah et al. \cite{ullah2018}. Range profile predictions using the {scalar} ESP calculated using SRIM are plotted for both the channel and random directions.
}
\end{figure}

We simulated range profiles in Ni using the UTTM model with the parametrization by Caro et al. \cite{tamm_2019_eph1}. The MDRANGE simulation cell for Ni channeling ranges consisted of face-centered cubic (FCC) unit cells with a lattice constant of $3.52$ Å \cite{exp_lat_const}. Figure \ref{fig:Ni_prof} shows simulated range profiles of 25 keV Ni ions in Ni, along the \(\langle 111 \rangle\) channeling direction and along random trajectories. 
Along the \(\langle 111 \rangle\) direction, we also compare to a range profile obtained using scalar ESP values calculated with rt-TDDFT by Ullah et al.\cite{ullah2018}. 

{Scalar ESP data calculated with rt-TDDFT is only available in literature along the \(\langle 111 \rangle\) direction in Ni, and not for other crystallographic directions. The integral range profile along this direction lacks the pronounced drop off at the end of the range profile, which is characteristic of strongly channeling directions as seen for principal lattice directions in Fig. \ref{fig:Si_ranges}. Nevertheless, the UTTM model gives a range profile consistent with that predicted using SRIM along random trajectories in a polycrystalline target.}

\subsection{\label{subsec:GaAs_ranges} Gallium Arsenide}

For simulations of the arsenic (As) range profile in a GaAs target, we generated a GaAs lattice with a lattice constant of $5.65$ Å \cite{exp_lat_const}. We simulated range profiles along the \(\langle 100 \rangle\), \(\langle 111 \rangle\), and \(\langle 110 \rangle\) channeling directions using the UTTM model with the multi-element parametrization developed by Teunissen et al., and compare to ranges simulated with scalar rt-TDDFT ESP values \cite{teunissen2023}. 

\begin{figure}
\includegraphics[width=0.49\textwidth]{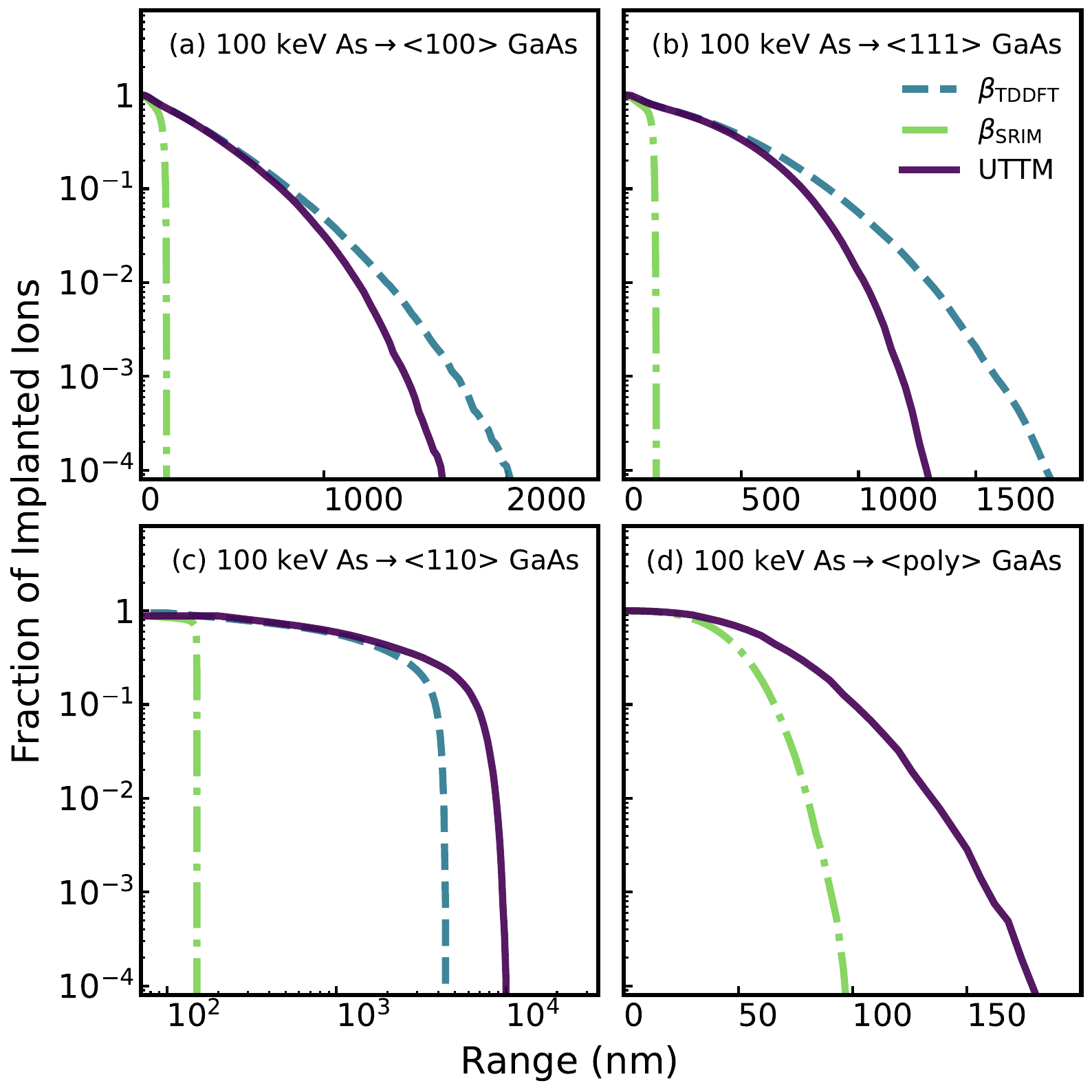}
\caption{\label{fig:gaas} Calculated integral range profiles of 100 keV As ions in DIA GaAs along the \(\langle 110 \rangle\),  \(\langle 100 \rangle\), and \(\langle 111 \rangle\) principal channeling directions, and along a random direction. The range profiles were simulated using scalar ESP values, which were calculated using TDDFT along channeling directions, and calculated by SRIM for the random direction. We compare these predictions with those obtained using the UTTM model for the ESP. The coupling function for the UTTM model was developed by Teunissen et al. \cite{teunissen2023}. Note the different axes scales, particularly the logarithmic x-axis scale in (c). }
\end{figure}

The range profile results are shown in Fig. \ref{fig:gaas}. The predictions along the \(\langle 110 \rangle\) channeling direction once again exhibit the deepest ion penetration depths with every ESP model, as was the case with Si, which has the same lattice type. The range profile prediction along the random direction using the UTTM model is not in agreement with that obtained using SRIM, highlighting the necessity of a robust parametrization to accurately capture trajectory-dependence of the ESP. 

\section{\label{sec:conc} Conclusions}

We presented the implementation of the trajectory-dependent UTTM model for the ESP in the efficient MD-based code MDRANGE. We simulated ion ranges for a variety of ion-target combinations and demonstrated that the UTTM model yields a clearly trajectory-dependent effective ESP. 
This implementation enables the efficient statistical calculation of range profiles based on thousands of trajectories that describes electronic energy losses to \textit{ab initio} accuracy. {These simulations can serve as a benchmark for validation and improvement in the development of $\alpha (\bar{\rho})$ fittings for the UTTM model.} 

These simulations {also} offer a route to validate the $\alpha (\bar{\rho})$ fitting against experimental measurements of ion ranges, which are much easier to obtain than direct measurements of the ESP \cite{eriksson1967}. 
Due to its computational efficiency, MDRANGE with the implementation of the UTTM model is also well-suited for direct comparison to ion transmission experiments, which measure energy losses from millions of ion trajectories through thin foils. This offers a route to directly benchmark the trajectory-dependent energy losses predicted by a parametrization of the UTTM model against experimentally measured values, and will be the focus of future work.

\begin{acknowledgments}
The authors wish to thank Alfredo A. Correa, Artur Tamm and Kai Nordlund for valuable discussions. This work has been partly carried out within the framework of the EUROfusion Consortium, funded by the European Union via the Euratom Research and Training Programme (Grant Agreement No. 101052200-EUROfusion). Part of this work was funded by the European Research Council (ERC, MUST, 101077454). Bruno Semião acknowledges support from the Aalto Science Institute of the Aalto University School of Science. Views and opinions expressed are, however, those of the author(s) only and do not necessarily reflect those of the European Union or the European Research Council. Neither the European Union nor the granting authority can be held responsible for them. 
\end{acknowledgments} 

\section*{\label{sec:app} Data Availability}

The source code for MDRANGE is available in a public Gitlab repository \cite{mdrange_link}. Example simulations using the UTTM model can be found in the repository.

Other data will be made available on request.

\bibliographystyle{IEEEtran}
\bibliography{bibliography} 

\end{document}